\documentclass[12pt]{article}

\usepackage{amssymb,amsmath,amsthm}
\usepackage{array,longtable}
\usepackage[mathscr]{eucal}

\usepackage[dvips]{graphicx}
\usepackage[dvips]{color}

\newcolumntype{V}{>{$}m{4cm}<{$}}
\newcolumntype{C}{>{$}c<{$}}
\newcolumntype{L}{>{$}l<{$}}
\newcolumntype{R}{>{$}r<{$}}

\newcommand{\ra}{\rightarrow}

\newcommand{\Ac}{\mathcal{A}}
\newcommand{\Bc}{\mathcal{B}}
\newcommand{\Ic}{\mathcal{I}}

\newcommand{\Qc}{\mathcal{Q}}

\newcommand{\Uc}{\mathcal{U}}
\newcommand{\Oc}{\mathcal{O}}
\newcommand{\pd}{\partial}

\newcommand{\Tr}{\mathop{\mathrm{Tr}}\nolimits}
\newcommand{\la}{\langle\!\langle}
\renewcommand{\ra}{\rangle\!\rangle}

\allowdisplaybreaks[3]

\begin{document}
\begin{center}
{\LARGE Solutions of Vacuum Superstring Field
Theory}\\
Alexey Koshelev\\
\textit{\small{Steklov Mathematical Institute, Moscow, Russia}}\\
\textit{\small{Physics Department, University of Crete, Greece}}\\
\texttt{\small{alex-kas@yandex.ru}}
\end{center}

\centerline{\textbf{Abstract}} 
In this report we
review\footnote{This report is based on the papers
\cite{ia_vssft}-\cite{ia_nsgs}.} a structure of cubic Vacuum
Superstring Field Theory and known solutions to its equation of
motion.
\section{Introduction}

During the last two years the bosonic vacuum string field theory (VSFT)
 proposed to describe physics around  the bosonic tachyon vacuum
  \cite{F2} has been  investigated in many papers
\cite{zwiebach}-\cite{Ohmori}.
VSFT action has the same form as
the original Witten SFT action \cite{Witten},
but with a new differential operator $\mathcal{Q}$ (for a review of
SFT see \cite{0102085,0109182,ABGKM} and references therein). The
absence of physical open string excitations around the tachyon
vacuum \cite{sen-con,Hata1}
supports a suggestion \cite{F2} that after some field
redefinition $\mathcal{Q}$ can be written as a pure ghost
operator.  Under this assumption
solutions to VSFT equation of motion
admit a  factorized form with the projector-like matter part.

A generalization of VSFT
to superstrings has been discussed in \cite{F2} and more recently in
\cite{ia_vssft}-\cite{ia_nsgs},\cite{Ohmori}
and \cite{Marino} in the context of cubic SSFT \cite{PTY,AMZ1} and
non-polynomial SSFT \cite{0002211},
respectively.
Open fermionic string in the NSR formalizm has a
tachyon in the GSO$-$ sector that leads to a classical instability of
the perturbative vacuum in the theory without supersymmetry. It
has been proposed  \cite{sen-con} to interpret the tachyon
condensation in the GSO$-$ sector of the NS string as a decay of unstable
non-BPS D9-brane.

In this note we consider a construction of cubic Vacuum Superstring field Theory
and solution to its equation of motion.
Actually, this means a construction of a new BRST charge while the structure of the action
will be the same. This question is considered in Section 2. In Section 3 we consider
solution to the matter part of the fermionic sector of the NS string, the NS sliver.
In Section 4 NS ghost sliver is considered.

\section{Cubic Vacuum String Field Theory on a non-BPS $\mathrm{D}$-brane }

To describe the open string states living on a
single non-BPS $\mathrm{D}$-brane one has to consider
GSO$\pm$ states \cite{sen-con}.
GSO$-$ states are Grassmann even, while
GSO$+$ states are Grassmann odd.
The unique (up to rescaling of the fields)
gauge invariant cubic action unifying GSO$+$
and GSO$-$ sectors is \cite{ABKM}
\begin{equation}
\begin{split}
S[A_+,A_-]&=\frac{1}{g^2_o}\left[
\frac{1}{2}\langle\!\langle Y_{-2}|A_+,Q_BA_+
\rangle\!\rangle+\frac{1}{3}\langle\!\langle
Y_{-2}|A_+,A_+,A_+\rangle\!\rangle
\right.\\
&~~~~~~~~~\left.+\frac{1}{2}\langle\!\langle
Y_{-2}|A_-,Q_BA_-\rangle\!\rangle
-\langle\!\langle
Y_{-2}|A_+,A_-,A_-\rangle\!\rangle\right].
\end{split}
\label{action7}
\end{equation}

Here the factors before the odd brackets are fixed by the
constraint of gauge invariance, that is specified below, and
reality of the string fields $A_{\pm}$. Variation of
this action with respect to $A_+$, $A_-$
yields the following equations of motion (see \cite{ABKM} for details)
\begin{equation}
Q_BA_{\pm}+A_+\star A_{\pm}
-A_-\star A_{\mp}=0
\label{eqmotion}
\end{equation}
The action \eqref{action7} is invariant under the
gauge transformations
\begin{equation*}
\delta A_{\pm}=Q_B\Lambda_{\pm}+[A_{\pm},\Lambda_+]
+\{A_{\mp},\Lambda_-\}
\end{equation*}
where $[\,,]$ ($\{\,,\}$) denotes $\star$-(anti)commutator
and $\Lambda_{\pm}$ are gauge parameters.

The action \eqref{action7}
can be rewritten in the matrix form
 as
\begin{equation}
S[\hat{A}]=\frac{1}{2g_o^2}\Tr\left[\frac{1}{2}\la Y_{-2}|\hat{A},\hat{Q}_B\hat{A}\ra
+\frac13\la Y_{-2}|\hat{A},\hat{A},\hat{A}\ra\right],
\label{theAction}
\end{equation}
$\hat{Q}_B=Q_B\otimes a,~\hat{Y}_{-2}=Y_{-2}\otimes a,~
\hat{A}=A_+\otimes a+A_-\otimes b
$ and $a$ and $b$ are $2\times 2$ matrices such that
$
a^2=1,~b^2=-1\text{ and }\{a,b\}=0.
$

The action \eqref{theAction} is invariant under GSO symmetry transformation given by
$\hat{A}\mapsto ((-1)^F\otimes 1)\hat{A}$,
and twist symmetry transformation $\Omega$ which action on the string field
is given via conformal transformation $M(z)=e^{-\pi i}z$.
One can check that the BRST charge $\hat{Q}_B$ commutes with $(-1)^F\otimes 1$
and $\Omega$.

Let $\hat{A}_{0}$ be a solution to the equations (\ref{eqmotion}).
A shift of a string field $
\hat{A}=\hat{A}_0+\hat{\Ac}
$ yields the following form of the action
\eqref{theAction}
\begin{equation}
S[\hat{A}_0,\hat{\Ac}]=S[\hat{A}_0]+\frac{1}{2g_o^2}\Tr\left[\frac{1}{2}\la Y_{-2}|\hat{\Ac}, \hat{Q}\hat{\Ac}\ra
+\frac13\la Y_{-2}|\hat{\Ac},\hat{\Ac}, \hat{\Ac}\ra\right],
\label{VSFTaction}
\end{equation}
where $\hat{Q}$ is ``a new BRST charge'' of the form
\begin{equation}
\hat{Q}=\hat{Q}_B+\{\hat{A_0},\cdot\}.
\label{shift:BRST}
\end{equation}
Further we will refer to $\hat{Q}$ as a kinetic operator.
One can check that the equation $\hat{Q}^2=0$
yields the equation of motion for the field $\hat{A_{0}}$ and
therefore $\hat{Q}$ is nilpotent.

The kinetic operator can be written in the form
\begin{equation}
\hat{Q}=Q_{\textsf{odd}}\otimes a+Q_{\textsf{even}}\otimes b.
\label{abg:100}
\end{equation}
The nilpotency of the $\hat{Q}$
yields the following identities for the operators $Q_{\textsf{odd}}$
and $Q_{\textsf{even}}$
\begin{equation}
Q_{\textsf{odd}}^2-Q_{\textsf{even}}^2=0\quad\text{and}\quad
[Q_{\textsf{odd}},\,Q_{\textsf{even}}]=0.
\label{eqBRST}
\end{equation}

Equations of motion following from the VSFT action \eqref{VSFTaction}
have the same form as for the action \eqref{theAction} but with the shifted BRST operator $\hat{Q}$.
In components these equations are
\begin{equation}
Q_{\textsf{odd}}\Ac_{\pm}-Q_{\textsf{even}}\Ac_{\mp}+\Ac_+\star \Ac_{\pm}
-\Ac_-\star \Ac_{\mp}=0.
\label{eqmotion-cal'}
\end{equation}

According to Sen conjectures \cite{sen-con} the solution $\hat{A}_0$
represents the
vacuum without open string excitations\footnote{This conjecture has been checked
for the non-BPS brane decay only at the first non-trivial
level {\cite{ABGKM}}.}, and therefore the cohomology of the kinetic
operator $\hat{Q}$ must be zero.

In proposing a simple form of the vacuum SSFT action, we have in mind
field redefinition, which preserves the form of the cubic action,
but simplifies the expression for the kinetic operator $\hat{Q}$.
By an appropriate field redefinition
\begin{subequations}
\begin{equation}
\hat{\Uc}=\Uc_{\textsf{even}}\otimes 1+\Uc_{\textsf{odd}}\otimes ab
\end{equation}
we will assume a $\star$-algebra homomorphism
$
\hat{\Uc}(\hat{\Ac}\star\hat{\Bc})=(\hat{\Uc}\hat{\Ac})\star(\hat{\Uc}\hat{\Bc}),
$
which satisfies two additional conditions:
\begin{equation}
\Tr\int'\hat{\Uc}\hat{\Ac}=\Tr\int'\hat{\Ac}
\text{ and an existance of }\hat{\Uc}^{-1}:\hat{\Uc}\hat{\Uc}^{-1}=1.
\end{equation}
\label{def:redef}
\end{subequations}
The  $~\hat{}~$  in the expressions for the field redefinition $\hat{\Uc}$
is very important since this transformation acts in both GSO$+$ and GSO$-$ sectors.
Using \eqref{def:redef} one can check that after the field redefinition
$
\hat{\Ac}\mapsto \hat{\Uc}\hat{\Ac}
$
the kinetic operator transforms into
$
\hat{\Qc}=\hat{\Uc}^{-1}\hat{Q}\hat{\Uc}.
$
Note that the transformation $\hat{\Uc}$ is highly
non-trivial and mixes GSO$+$ and GSO$-$
sectors.

Consider the standard BRST charge in the
superconformal field theory
\begin{equation}
Q_B=\frac{1}{2\pi i}\oint d\zeta \Bigl[
c(T_B+T_{\phi}+T_{\eta\xi}+\frac{1}{2}T_{bc})
-\eta e^{\phi}T_F+\frac{1}{4}b\pd\eta\eta e^{2\phi}
\Bigr].
\label{BRST:std}
\end{equation}
One can check that after the homogeneous field redefinition
\cite{9902178}
\begin{equation}
\Uc=e^{-R},\quad\text{where}\quad
R=\frac{1}{2\pi i}\oint d\zeta \Bigl[
cT_F e^{-\phi}e^{\chi}+\frac{1}{4}\pd(e^{-2\phi})e^{2\chi}c\pd c
\Bigr]\label{redef-1}
\end{equation}
the BRST charge \eqref{BRST:std} takes the form
\begin{equation}
\Qc=\Uc^{-1}Q_B\Uc=\frac{1}{2\pi i}\oint d\zeta\,
b\gamma^2(\zeta).\label{redef-2}
\end{equation}

Following the idea of the paper \cite{F2},
which is based on Sen conjectures, gauge invariance
and algebraic properties of the BRST charge,
we require $\hat{\Qc}$ to satisfy the following properties:
\begin{enumerate}
\item $\hat{\Qc}=\Qc_{\textsf{odd}}\otimes a + \Qc_{\textsf{even}}\otimes b$;
\item Both $\Qc_{\textsf{odd}}$ and $\Qc_{\textsf{even}}$ have superghost number
equal to one, but $\Qc_{\textsf{odd}}$ is Grassmann odd, while
$\Qc_{\textsf{even}}$ is Grassmann even;
\item $\hat{\Qc}$ is a nilpotent operator, that in components means
the identities
$$
\Qc_{\textsf{odd}}^2-\Qc_{\textsf{even}}^2=0
\quad\text{and}\quad
[\Qc_{\textsf{odd}},\Qc_{\textsf{even}}]=0;
\label{Qequations}
$$
\item
$
\hat{\Qc}(\hat{A}\star\hat{B})=(\hat{\Qc}\hat{A})\star\hat{B}+(-1)^{|\hat{A}|}
\hat{A}\star(\hat{\Qc}\hat{B}),
$
In particular, this identity means that operators $\Qc_{\textsf{odd}}$ and $\Qc_{\textsf{even}}$
also satisfy the Leibnitz rule;
\item
$
\Tr \int'\hat{\Qc}(\hat{\Ac}\star\hat{\Bc})=0;
$
\item The operator $\hat{\Qc}$ must be universal, what means
that it has to be written without reference to the brane
boundary CFT;
\item The operator $\hat{\Qc}$ must have vanishing cohomology;
\item
$
[\hat{Y}_{-2},\,\hat{\Qc}]=0\quad\text{or}\quad
\{\hat{Y}_{-2},\,\hat{\Qc}\}=0.
$
We need this axiom to relate the axiom~5 with the fact
that $\hat{\Qc}$ annihilates the identity $|\Ic\rangle$. Therefore
we can have several variations of this axiom and in general
we only need something like the following
\begin{equation*}
\Qc_{\textsf{odd}}Y_{-2}\pm Y_{-2}\Qc_{\textsf{odd}}=0\quad\text{and}\quad
\Qc_{\textsf{even}}Y_{-2}\pm Y_{-2}\Qc_{\textsf{even}}=0;
\end{equation*}
Plus/minus in these formulae can be chosen independently.
\item $\hat{\Qc}$ is a hermitian operator, which means that both
$Q_{\textsf{odd}}$ and $Q_{\textsf{even}}$ are hermitian ones.
\end{enumerate}

Since $A_{0,+}\neq 0$ and $A_{0,-}\neq 0$ we believe that after the field
redefinition both charges $\Qc_{\textsf{odd}}$ and
$\Qc_{\textsf{even}}$ are non zero.

The following ghost kinetic operator satisfies all above axioms
\cite{ia_vssft,Ohmori}
\begin{subequations}
\begin{align}
\Qc_{\textsf{odd}}&=\frac{\mu^2}{4i}\,\bigl[c(i)-c(-i)\bigr]
+\frac{1}{2\pi i}\oint b(z)\gamma^2(z)dz,
\\
\Qc^{+}_{\textsf{even}}&=\frac{\mu}{2i}\,\bigl[\gamma(i)-\gamma(-i)\bigr],~
\Qc^{-}_{\textsf{even}}=\frac{\mu}{2}\,\bigl[\gamma(i)+\gamma(-i)\bigr],
\end{align}
\label{formal-q}
\end{subequations}
where  $\Qc^{\pm}_{\textsf{even}}$ means the restriction of the
operator $\Qc_{\textsf{even}}$ to GSO$\pm$ sectors and $\mu$ is a complex number.

In some sense
\eqref{formal-q} is the only form for the kinetic operator which
satisfies the twist invariance and the above conditions. One can explain it as follows. Following
\cite{0111129} consider an original (before field redefinition) BRST
charge $Q$ defined as
\begin{equation}
Q=\sum_{r}\frac{1}{2\pi i}\oint d\zeta\, a_{r}(\zeta)\Oc_{r}(\zeta)
\label{ser}
\end{equation}
where $a_{r}$ are smooth forms of $\zeta$ and $\Oc_{r}(\zeta)$ are some local
conformal operators of ghost number $1$. It was shown in \cite{0111129} that
after a singular field redefinition
the  dominant contribution to the transformed charge $\Qc$
will come from the lowest dimensional conformal operators.
This has led to the choice of $c(i)$ and $c(-i)$ in the bosonic case,
and this also leads to our choice of $\Qc_{\textsf{even}}$, since $\gamma$ is
the lowest dimensional even primary operator of ghost number $1$.


\section{NS Matter Sliver}

While after the field redefinition the kinetic operator of VSFT
has a pure ghost form it is natural to search for solutions to
VSFT equation of motion
in the factorized form $\Phi =\Xi_{matter}\otimes \Phi_{ghost}$,
where $\Xi_{matter}$ satisfies a
projector-like equation:
\begin{equation}
\Xi_{matter}=\Xi_{matter}\star~\Xi_{matter}. \label{proj}
\end{equation}
An equation similar to (\ref{proj}) has appeared in a construction
of solitonic solutions in noncommutative field theories in the
large non-commutativity limit \cite{GMS}.

A way to solve projection equation (\ref{proj}) for
the bosonic matter has been proposed by Rastelli and Zwiebach
\cite{zwiebach}. They have constructed a solution to (\ref{proj})
as the $n\rightarrow\infty$ limit of the wedge states $|n\rangle$.
The wedge states are defined on CFT language and they satisfy the
algebra
\begin{equation}
|n\rangle\star |m\rangle=|n+m-1\rangle. \label{wedge-al}
\end{equation}
From algebra (\ref{wedge-al}) it immediately follows that $|\infty\rangle$,
the so-called sliver state, satisfies (\ref{proj}).

Now we are going to construct the fermionic sliver state using CFT methods.
We refer reader to \cite{ia_nsms} in order to find the algebraic construction
of the fermionic sliver state. We have to note that numeric calculations show
a conspicous agreement between algebraic and CFT methods \cite{ia_nsms}.

A generalization of the bosonic wedge states
\cite{zwiebach,RSZF} to the fermionic wedge states is
straightforward.
Wedge states $|n\rangle$ are defined by
\begin{equation}
\langle n|\phi^{\psi}\rangle=\langle f_{n}\circ\phi^{\psi}(0)\rangle,
\end{equation}
where $|\phi^{\psi}\rangle$ is an arbitrary state which belongs to
the fermionic subspace, $f_{n}\circ\phi^{\psi}(\xi)$ denotes the
conformal transform of $\phi^{\psi}(\xi)$ and $f_{n}(\xi)$ is
the same as in the bosonic case, i.e.
\begin{gather}
f_{n}(\xi)=\frac{n}{2}\tan\left(\frac{2}{n}\tan^{-1}\xi\right).\label{wedge-map}
\end{gather}
The wedge state with $n=1$ corresponds to the identity of the star
algebra and with $n=2$ corresponds to the vacuum.

 Taking the limit
$n\rightarrow\infty$ in \eqref{wedge-map} one derives the
conformal map for the sliver state $|\infty\rangle$
\begin{equation}
w(\xi)=\tan^{-1}(\xi).\label{sliv-map}
\end{equation}
For a state
$|\Lambda\rangle\propto\exp(1/2\psi^{\dag}_{r}\Lambda_{rs}\psi^{\dag}_{s})|0\rangle,
$
corresponding to a
conformal map $\lambda(\xi)$ one gets
\begin{equation}
\Lambda_{rs}=\oint\frac{d\xi}{2\pi\imath}\oint\frac{d\xi'}{2\pi\imath}\xi^{-r-
\frac{1}{2}}\xi'{}^{-s-\frac{1}{2}}
\left(\frac{\partial
\lambda(\xi)}{\partial\xi}\right)^{\frac{1}{2}}\frac{1}{\lambda(\xi)-\lambda(\xi')}\left(\frac{\partial
\lambda(\xi')}{\partial\xi'}\right)^{\frac{1}{2}}.\label{state-map}
\end{equation}
Here $\oint$ denotes the contour integration around the origin.
Substituting the sliver conformal map \eqref{sliv-map} one gets that the conformal sliver
$|\tilde{\Xi}^{\psi}\rangle\equiv|\infty\rangle$ is defined as
\begin{equation}
|\tilde{\Xi}^{\psi}\rangle=\tilde{\mathcal{N}}^{10}\exp(\frac{1}{2}\psi^{\dag}_{r}\tilde{S}_{rs}\psi^{\dag}_{s})|0\rangle,
\end{equation}
\begin{equation}
\tilde{S}_{rs}=\oint\frac{d\xi}{2\pi\imath}\frac{d\xi'}{2\pi\imath}\xi^{-r-\frac{1}{2}}\xi'{}^{-s-\frac{1}{2}}
\frac{2\imath}{\sqrt{1+\xi^2}\,\sqrt{1+\xi'{}^2}}\ln\left(\frac{(1+i\xi)(1-i\xi')}{(1-i\xi)(1+i\xi')}\right).
\end{equation}
The matrix $\tilde{S}_{rs}$ can be calculated explicitly. Only
coefficients with $r+s=\text{even}$ differ from zero.


\section {NS ghost sliver}

Ghost part of VSFT equations of motion has been studied in
\cite{0108150,0111129}.
 It was observed that a sliver constructed in the twisted conformal theory
with new $SL(2,\mathbb{R})$ invariant vacuum
solves the ghost part of VSFT  equation of motion.
This equation is a usual SFT  equation of motion with a canonical choice
of ghost kinetic term
that is a local insertion at the  string midpoint.

We present here the twisted superghost conformal theory
and derive corresponding equations in analogy with the one
constructed by Gaiotto, Rastelli, Sen and
Zwiebach \cite{0111129}. We refer the reader to \cite{ia_nsgs} where algebraic construction
of the NS ghost sliver can be found.

A twisted CFT is introduced by subtracting from the stress energy tensor $T(w)$
of the $(\beta,\gamma)$ system the derivative of $U(1)$ ghost number current
$j$ as follows
\begin{gather}
T'(w)=T(w)-\pd j(w),\quad \Bar{T}'(\Bar{w})=\Bar{T}(\Bar{w})-\pd\Bar{j}(\Bar{w}),\quad j=-\beta\gamma.
\end{gather}
More explicitly for the holomorphic stress energy tensor one obtains
\begin{eqnarray}
T(w)=-\frac32\beta\pd\gamma(w)-\frac12\pd\beta\gamma(w),\quad \mathrm{with~}c=11,\\
T'(w)=-\frac12\beta'\pd\gamma'(w)+\frac12\pd\beta'\gamma'(w),\quad
\mathrm{with~}c=-1,
\end{eqnarray}
where $(\beta',\gamma')$ denotes the superghosts of the twisted CFT and $c$ is the central charge.
The weights of these new $\beta'$ and $\gamma'$ become equal
to $1/2$
and the superghost current $j'=-\beta'\gamma'$ has no anomaly.
Fermionic ghosts in the original theory are bosonised as
\begin{gather}
\gamma(w)=\eta e^{\phi}(w),\quad
\beta(w)= e^{-\phi}\pd\xi(w),
\end{gather}
so that the ghost number current is expressed in the form $j=-\pd\phi$.
The Euclidean world-sheet actions
$S$ and $S'$ for the fields $\phi$ and $\phi'$ correspondingly are related as
\begin{gather}
S[\phi]= S'[\phi]-\frac{1}{2\pi}\int_{\Sigma}
d^{2}\zeta\;\sqrt{g}R^{(2)}(\phi+\Bar{\phi}),
\end{gather}
where $\zeta$ denotes the world-sheet coordinates, $g$ denotes the
Euclidean world-sheet metric and $R^{(2)}$ is the scalar
curvature.

We assume that scalar curvature is proportional to $\delta$-function,
which has a support on the infinity in some coordinates on $\Sigma$.
Therefore we can identify the fields $\phi$ and $\phi'$ of two CFTs.
The states in the two theories can be identified by
the following map between the oscillators and the vacuum states
\begin{gather}
\beta_{n}\leftrightarrow\beta'{}_{n},\quad \gamma_{n}\leftrightarrow\gamma'{}_{n},\quad
|-1\rangle\leftrightarrow|0'\rangle,\quad \langle
-1|\leftrightarrow\langle 0'|,\quad
\langle0'|0'\rangle=1,\label{cooresp}
\end{gather}
where $|0\rangle$ and $|0'\rangle$ are the $SL(2,\mathbb{R})$
invariant vacua  of two  theories and $|-1\rangle=e^{-\phi(0)}|0\rangle$.

In the CFT$'$ the fields $\beta',\gamma'$ are bosonized as in the original
theory
\begin{gather}
\gamma'(w)=\eta e^{\phi}(w),\quad
\beta'(w)= e^{-\phi}\pd\xi(w).
\end{gather}
Notice that we do not introduce new notations for the
$(\xi,\eta)$ system because it has  not changed.

The advantage of the CFT method in comparison with the operator method,
that we have used in  Section 2, is that we do not have
to postulate the sliver equation from the very beginning.
The aim of this section is to define a sliver state as a surface state
over $SL(2,\mathbb{R})$ invariant vacuum in CFT and  CFT$'$, correspondingly,
by the conformal map used in the matter case.

First we define the surface state for the original $(\beta,\gamma)$ system.
The fermionic ghost surface state
corresponding to the conformal map $\lambda(\xi)$ is defined as
\begin{gather}
\langle\Lambda|=\mathcal{N}_{\beta\gamma} \langle 0|\exp(-\sum_{r\geq 3/2\atop s\geq -1/2}\gamma_{r}\Lambda_{rs}\beta_{s}),
\end{gather}
where $\mathcal{N}_{\beta\gamma}$ is a normalization factor and
the matrix $\Lambda_{rs}$ is defined so that
the following identity holds
\begin{gather}
\langle 0|e^{-\sum\limits_{r\geq 3/2 \atop s\geq
-1/2}\gamma_{r}\Lambda_{rs}\beta_{s}}\gamma(w)
\beta(z)e^{-Q\phi(0)}|0\rangle=
\langle \lambda\circ \gamma(w)\lambda\circ
\beta(z)\lambda\circ e^{-Q\phi(0)}\rangle.
\label{def-state-map-h-2}
\end{gather}

One can evaluate $\Lambda_{rs}$
explicitly. To this end one has to calculate the left hand side and
right hand side of \eqref{def-state-map-h-2}.
Substitution of
$\gamma(w)=\sum_{r}\gamma_{-r}w^{r+1/2}$ and
$\beta(z)=\sum_{s}\beta_{-s}z^{s-3/2}$ into the left hand side of \eqref{def-state-map-h-2}
yields
\begin{gather}
h(z,w)\equiv\langle 0|e^{-\gamma_{r}\Lambda_{rs}\beta_{s}}\gamma(w)\beta(z)e^{-Q\phi(0)}|0\rangle= -\sum_{r,s}w^{r+1/2}z^{s-3/2}\Lambda_{rs},\end{gather}
therefore
\begin{gather}
\Lambda_{rs}=-\oint\frac{dz}{2\pi i}\frac{1}{z^{r-1/2}}
\oint\frac{dw}{2\pi i}\frac{1}{w^{s+3/2}}h(z,w).
\end{gather}

Further one evaluates the correlation function in the right hand side of
\eqref{def-state-map-h-2}
\begin{multline}
\langle \lambda\;\circ\; \gamma(w)\lambda\;\circ\; \beta(z)\lambda\;\circ\; e^{-Q\phi(0)}\rangle\\
=\left(\frac{\pd\lambda(w)}{\pd w}\right)^{-1/2}
\left(\frac{\pd\lambda(z)}{\pd z}\right)^{3/2}
\frac{1}{\lambda(w)-\lambda(z)}\left(\frac{\lambda(w)-\lambda(0)}{\lambda(z)-\lambda(0)}\right)^{-Q}.
\end{multline}
One gets the following answer for $\Lambda_{rs}$
\begin{equation}
\oint\frac{dz}{2\pi i}\frac{z^{\frac12}}{z^{r}}
\frac{dw}{2\pi i}\frac{w^{-\frac32}}{w^{s}}
\left(\frac{\pd\lambda(w)}{\pd w}\right)^{-\frac12}\!\!\!
\left(\frac{\pd\lambda(z)}{\pd z}\right)^{\frac32}\!\!
\frac{1}{\lambda(z)-\lambda(w)}
\left(\frac{\lambda(z)-\lambda(0)}{\lambda(w)-\lambda(0)}\right)^2.
\label{lambda-conf-sliver0}
\end{equation}

The fermionic ghost surface state in CFT'
corresponding to the conformal map $\lambda(\xi)$ is defined as
\begin{gather}
\langle\Lambda'|=\mathcal{N}'{}_{\beta\gamma} \langle 0'|\exp(-\sum_{r\geq 1/2\atop s\geq 1/2}\gamma_{r}\Lambda'{}_{rs}\beta_{s}),
\end{gather}
where $\mathcal{N}'{}_{\beta\gamma}$ is a normalization factor and
the matrix $\Lambda'{}_{rs}$ is defined so that
the following identity holds
\begin{gather}
\langle 0'|\exp(-\sum_{r\geq 1/2 \atop s\geq
1/2}\gamma_{r}\Lambda'{}_{rs}\beta_{s})\gamma'(w)
\beta'(z)|0'\rangle=
\langle \lambda\circ \gamma'(w)\lambda\circ
\beta'(z)\rangle'.
\label{def-state-map-h-2-1}
\end{gather}

Substitution of
$\gamma'(w)=\sum_{r}\gamma_{-r}w^{r-1/2}$ and
$\beta'(z)=\sum_{s}\beta_{-s}z^{s-1/2}$ into the left hand side of \eqref{def-state-map-h-2-1}
 yields
\begin{gather}
h'(z,w)\equiv\langle 0'|e^{-\gamma_{r}\Lambda'{}_{rs}\beta_{s}}\gamma'(w)
\beta'(z)|0'\rangle=-\sum_{r,s}w^{r-1/2}z^{s-1/2}\Lambda'{}_{rs},
\label{hminusone-1}
\end{gather}
therefore
\begin{gather}
\Lambda'{}_{rs}=-\oint\frac{dz}{2\pi i}\frac{1}{z^{r+1/2}}
\oint\frac{dw}{2\pi i}\frac{1}{w^{s+1/2}}h'(z,w).
\end{gather}

Evaluating the correlation function in the right hand side of
\eqref{def-state-map-h-2-1} one finds
\begin{equation}
\langle \lambda\;\circ\; \gamma'(w)\lambda\;\circ\; \beta'(z) \rangle'
=\left(\frac{\pd\lambda(w)}{\pd w}\right)^{1/2}
\left(\frac{\pd\lambda(z)}{\pd z}\right)^{1/2}
\frac{1}{\lambda(w)-\lambda(z)}.
\end{equation}
One gets the following answer
\begin{gather}
\Lambda'{}_{rs}=\oint\frac{dz}{2\pi i}\frac{1}{z^{r+1/2}}
\frac{dw}{2\pi i}\frac{1}{w^{s+1/2}}
\left(\frac{\pd\lambda(z)}{\pd z}\right)^{1/2}
\left(\frac{\pd\lambda(w)}{\pd w}\right)^{1/2}
\frac{1}{\lambda(z)-\lambda(w)}.\label{lambda-conf-sliver-1}
\end{gather}

It should be mentioned here that the matrix
\eqref{lambda-conf-sliver-1} coincides with the matrix of the matter
sliver.

\section*{Acknowledgments}

The author would like to thank I.Ya.~Arefeva, D.M.~Belov, A.A.~Giryavets and
P.B.~Medvedev for many valuable discussions and comments.
 This work was supported in part by RFBR grant
02-01-00695, RFBR grant for leading scientific schools, by
INTAS grant 99-0590 and by INTAS fellowship for Young Scientists YSF~2002-42.


\end{document}